\documentclass[12pt,preprint]{aastex}

\shorttitle{Molecular Emission Line Formation in Prestellar Cores}
\shortauthors{Pavlyuchenkov et al.}

\begin{document}
\title{Molecular Emission Line Formation in Prestellar Cores}

\author{
Ya. Pavlyuchenkov\altaffilmark{1,2}, 
D. Wiebe\altaffilmark{2}, 
B. Shustov\altaffilmark{2},}
\author{
Th. Henning\altaffilmark{1},
R. Launhardt\altaffilmark{1},
D. Semenov\altaffilmark{1}} 
\altaffiltext{1}{Max Planck Institut f\"ur Astronomie, 
K\"onigstuhl 17, D69117 Heidelberg, Germany;}
\altaffiltext{2}{Institute of Astronomy, Russian Academy of Sciences,
Pyatnitskaya 48, Moscow, 109117, Russia}


\begin{abstract}
We investigate general aspects of molecular line formation
under conditions which are typical of prestellar cores.  Focusing on
simple linear molecules, we study formation of their rotational lines by
radiative transfer simulations. We present a
thermalization diagram to show the effects of collisions and radiation
on the level excitation. We construct a detailed scheme (contribution
chart) to illustrate the formation of emission line profiles. This chart
can be used  as an efficient tool to identify which parts of the cloud
contribute  to a specific line profile. We show how molecular
line characteristics for uniform model clouds  depend on hydrogen
density, molecular column density, and kinetic temperature.  The results
are presented in a 2D plane to illustrate cooperative effects of the
physical factors.  We also use a core model with a non-uniform
density distribution and chemical stratification to study the effects of
cloud contraction and rotation on spectral line maps. We discuss the
main issues that should be taken into account when dealing with
interpretation and simulation of observed molecular lines.
\end{abstract}


\keywords{line: formation --- line: profiles ---
radiative transfer --- stars: formation --- ISM: clouds}

\section{Introduction}

Star formation is a fundamental process in the universe.  Dense,
gravitationally bound, and genuinely starless cores, which we call here 
``prestellar cores'', are the earliest visible precursors of forming
stars.  The interplay between gravitation and thermal/magnetic pressure
as well as the conservation of angular momentum are all driving
forces behind prestellar core evolution. While the thermal structure
is determined by dust properties and molecular composition, magnetic
support is sensitive to the ionization degree. The interaction and relative
importance of these processes as well as the role of external forces
are still not well understood \citep[see, e.g. reviews by][]{Klessen:2004, Bergin:2007}.
Thus, it is important to study the physical and chemical evolution of these cores
in detail to reveal the underlying physics.

The interiors of such  prestellar cores are well-shielded from
interstellar or stellar radiation, leading to low internal temperatures.
Under such conditions, the main component of these cores,  H$_2$, is not
easily observable. Therefore, we have to rely on indirect methods to
determine the physical structure of prestellar cores, e.g., on
observations of thermal dust emission or emission lines from
other molecules, like CO, CS and N$_2$H$^+$.  Observations of spectral
lines have an important advantage  over continuum observations since
they also carry information about kinematics of the gas. A disadvantage of
molecular tracers can be that they are only present under certain
conditions and can be frozen-out on dust surfaces. Numerous studies,
including single-dish and interferometric observations along with their
theoretical analysis, have been performed over the past years,
significantly deepening our understanding of the physical and chemical 
processes in star-forming regions 
\citep[see, e.g. reviews of][]{Myers:1999,Evans:1999,DiFrancesco:2007,
Bergin:2007}.  However, deriving physical properties from molecular
lines is a difficult (inverse) problem  because of the many factors which
can affect the line formation.

Even if we assume that the studied object is spherically symmetric and
uniform, we have to specify at least 5 independent parameters describing
the formation of molecular lines. These are density $n({\rm H}_2)$, kinetic
temperature $T_{\rm kin}$, molecular column density $N({\rm mol})$,
radial velocity $V_r$, and micro-turbulent velocity $V_{\rm turb}$. In
more realistic models, additional parameters should also be considered,
e.g., electron concentration and external radiation field, with spatial
distributions of all  the above parameters. 

An exact analytical treatment of this multi-parameter system is, in most
cases, impossible. Optically thin line formation can be described
analytically to some extent \citep[see, e.g.,][]{Wilson:2000},
but it is very difficult to do the same for optically thick lines. 
Also, it is difficult to use analytic methods to solve the inverse
problem, i.e., to restore physical distributions of relevant parameters
from observed spectra because of the non-local and non-linear nature of
the radiative transfer problem. This is why numerical line radiative
transfer (LRT) models are commonly used as an exploratory tool. Several
reliable numerical methods and numerical codes have been developed by
various groups for this purpose \citep[see reviews
by][]{Peraiah:2004,Zadelhoff:2002}.  There are also fast approximate
numerical tools available for the molecular line radiative transfer
analysis\footnote{{\it http://www.sron.rug.nl/\~{}vdtak/radex/radex.php}}
 \citep{vanderTak:2007}. However, input 
data for line
modeling include not only physical conditions but also distributions   
of molecular abundances. All these data can be represented with
some analytical prescription or extracted from chemical and dynamical
models.

As recently outlined by \citet{tsamis}, there are two alternative
approaches to the diagnostics of protostellar objects. The first
approach is a systematic study of the influence of various factors on
emergent spectra in order to facilitate the analysis of forthcoming
observations. Such studies (based mainly on approximate LRT methods)
have already started more than 30 years ago  with the analysis of
the excitation conditions for various molecules and effects of the
density, temperature and velocity gradients on the emission line
profiles 
\citep[e.g.,][]{Lucas:1974,Goldreich:1974,Leung:1978,Stenholm:1980}. 
The alternative approach is a detailed study of an individual
source. To infer
its parameters, a number of models is constructed and the direct LRT
problem is solved  for each model. Varying the model parameters, it is
possible to find  the combination which provides the best agreement
between observed and modeled spectra (maps) or their derived parameters,
like, e.g., velocity centroids \citep{walker1994} or line ratios
\citep{vanderTak:2007}. Both trial-and-error methods or any sophisticated
numerical algorithm for isolating the ``best''\  set of free parameters
can be utilized \citep{Keto:2004}. This second approach has already been
successfully applied in a number of studies, aimed to extract detailed
characteristics of pre-stellar and protostellar objects from observed
spectra, in particular, their chemical and kinematical structure
\citep[e.g.,][]{Tafalla:2002,Tafalla:2004,Keto:2004,
Evans:2005,Brinch:2007} as well as to test different star formation
theories \citep[e.g.][]{Pavlyuchenkov:2003, Offner:2007}.

Following the first approach, efforts of various authors  are mainly
concentrated toward evolutionary stages later than the prestellar phase. Most
cores investigated in detail are closer to the formation of a first
hydrostatic core, when the collapse  is well-developed and can be
described by Shu or Larson-Penston solutions.  This interest may be
partly caused by the fact that first proofs of infall have been reported
for Class\,0 objects \citep{walker1986,b335}, in which  the collapse has
already resulted in the formation of a deeply embedded source. 
\cite{zhou1992} considered spectral differences between Shu and 
Larson-Penston solutions, assuming constant molecular abundances, and
identified four line properties, supposedly unique to collapsing cores.
\citet{rawlings1992} combined a dynamical  description, based on the Shu
model, with a detailed chemical model and identified species  with broad
line wings, indicative of infall motions. \citet{ry2001} coupled a
similar chemical and dynamical model to a more realistic radiation
transfer model, implemented as an approximate $\Lambda$-iteration code,
and studied the sensitivity of emergent central optically thin and
optically thick spectra to various model parameters. 
The assumption of  isothermality, used by \citet{rawlings1992} and
\citet{ry2001}, was relaxed in \citet{tsamis} in application to a
``B335-like'' Class~0 object. The sensitivity of line  profiles in this
model to the intensity of the ambient radiation field was studied in 
\citet{redman2004}.
The influence of various parameters on Class~0 central spectra, with
a particular emphasis on turbulence and rotation, was also studied by
\citet{wtb2001} under the  assumption of flat abundance distributions.
The role of high angular resolution in studying such clouds was
investigated by \citet{choi}. More details about line modeling 
can be also found in review by \citet{Evans:1999}.


Large amounts of data have also been accumulated for starless cores, 
most of which are presumably ``prestellar'' \citep[e.g.,][]{infall},
together with the theoretical modeling of molecular lines. E.g.,
\citet{Tafalla:2004} investigated the effects of depletion on the line
intensity. \citet{Lee:2004} studied the evolution of line profiles
during core contraction based on the models where the chemical
evolution is calculated along with dynamical evolution of the cloud. 
\citet{Pavlyuchenkov:2007} investigated the combined effects of 
temperature and depletion, and \citet{DeVries:2005} investigated
the possibility to use approximate analytical models to extract velocity
gradients from the line profiles. In principle, all the mentioned studies of line
formation in protostellar objects are also relevant for prestellar cores.
However, despite the obvious
observational and theoretical progress, the interpretation of the data
is still far from being straightforward, as infall and rotation
velocities in prestellar cores are both comparable or even smaller than
the sound speed. Difficulties in restoring the information about
structural, thermal, and kinematic properties  are also caused by the
lack of relevant methods for molecular line analysis.

In this paper, we systematically study the formation of molecular lines 
in prestellar cores and present different tools which can be used to
analyze the formation of lines in detail. Using the linear molecules CO and
HCO$^+$ as examples, we model their emission by means of non-LTE LRT
simulations. The role of hydrogen density, molecular column density,
kinetic temperature, infall, and rotation is examined and illustrated
in the paper. Such an analysis may form the basis of a more
sophisticated interpretation of observational data and can be useful for
those who are going to use LRT simulations in their studies
or want to get a global view of the factors influencing observed
line profiles.

In Section~2 we describe the radiation transfer model used in the paper,
introduce parameters to characterize line profiles, and provide
some basic considerations on the line formation in prestellar cores. In
Section~3 we show how molecular line characteristics depend on hydrogen
density, molecular column density, and kinetic temperature for the
parameter sample of uniform model clouds. In Section~4 we consider a
non-uniform model cloud and study the effects of chemical
stratification, contraction and rotation on spectral line maps. In
Section~5 we discuss  additional problems related to the LRT analysis. 
Section~6 summarizes the main conclusions of this paper.

\section{Notes on Molecular Line Transfer}

\subsection{Equations of Radiation Transfer}


The goal of the LRT simulation is to obtain level populations and to
produce appropriate molecular line profiles. For that  purpose, one
solves a system of equations, describing the  radiative transfer
\citep[e.g., ][]{Peraiah:2004}. This system includes the  transfer
equation itself
\begin{equation}
\frac{dI_{\nu}}{ds}=-\alpha_{\nu}I_{\nu}+j_{\nu}
\label{transfer}
\end{equation}
and a set of balance equations for level populations
\begin{equation}
n_u \left[ \sum_{l<u}A_{ul} + \sum_{l\neq u}
(B_{ul}\overline{J}_{ul}+C_{ul})\right]= 
\sum_{l>u}n_{l}A_{lu}+\sum_{l\neq u}n_{l}(B_{lu}\overline{J}_{ul}+C_{lu}).
\label{balance}
\end{equation}
Here $I_{\nu}$ is the intensity of radiation, $s$ is the length along
the ray, $n_u$ and $n_l$ are level populations, $A_{ul}$ and $B_{ul}$
are the Einstein coefficients;  $C_{ul}$ are coefficients of collisional
excitation; $ul$ indices specify the transition from the upper level $u$
to the lower level $l$.  Equations (\ref{transfer}) and (\ref{balance})
are coupled by the emission and absorption coefficients
$j_{\nu}$ and $\alpha_{\nu}$, respectivelly: 
\begin{eqnarray}
&j_{\nu}=\frac{h\nu_{ul}}{4\pi}n_{u}A_{ul}\phi_{ul}(\nu), \\
&\alpha_{\nu}=\frac{h\nu_{ul}}{4\pi}(n_{l}B_{lu}-n_{u}B_{ul})\phi_{ul}(\nu)
\end{eqnarray}
and by the mean intensity $\overline{J}_{ul}$. The mean intensity
is defined as
\begin{equation}
\overline{J}_{ul}=\frac{1}{4\pi}\int\limits_{4\pi} d\Omega
\int\limits_0^{\infty} I_{\nu}\,\phi_{ul}(\nu) d\nu.
\end{equation}
where $\phi_{ul}(\nu)$ is the line profile function
and $\Omega$ is the spatial angle.
The line profile function can be expressed as
\begin{equation}
\phi_{ul}(\nu)=\frac{c}{b\nu_{ul} \sqrt{\pi}}\exp 
\left( -\frac{c^2(\nu-\nu_{ul} -
(\vec{v}\,\vec{n})\;\nu_{ul}/c)}{\nu_{ul}^2 b^2} \right)
\end{equation}
in the approximation of total redistribution over frequencies and a 
Maxwellian turbulent velocity distribution.
Here, $\nu_{ul}$ is the central frequency of 
the transition $u \rightarrow l$; $\vec{v}$ is the regular velocity; 
$\vec{n}$ is unit vector associated with $d\Omega$; and $b$ is a 
parameter, which is related to the kinetic temperature $T_{\rm kin}$ and the 
most probable value of the microturbulent velocity $V_{\rm t}$ by the 
expression
\begin{equation}
b^2=\sqrt{\frac{2kT_{\rm kin}}{m_{\rm mol}}+V_{\rm t}^2}.
\label{vturb}
\end{equation}
The intensity $I_{\nu}$ can be expressed in units of the
brightness temperature $T_{\rm B}$ via the Planck equation
\begin{equation}
I_{\nu}=\frac{2h\nu^3}{c^2}\frac{1}{e^{\frac{h\nu}{kT_{\rm B}}}-1},
\end{equation} 
or in the units of radiative temperature $T_{\rm R}$ via the
Rayleigh-Jeans approximation with subtracted background radiation
\begin{equation}
T_{\rm R}=\frac{c^2}{2k\nu^2}(I_{\nu}-I_{\nu}^{\rm bg}),
\end{equation} 
where $I_{\nu}^{\rm bg}$ is the intensity
of the background radiation with the temperature $T_{\rm bg}$. 
The radiative temperature $T_{\rm R}$ is commonly used in 
radio astronomy and we use it throughout this paper.

In order to calculate the emergent profile, one first calculates level populations 
and then integrates Eq.~(\ref{transfer}) to get a spectrum, which can
be later convolved with a telescope beam, if necessary. For our LRT simulations
we use the 2D non-LTE code URAN(IA), developed by \cite{Pavlyuchenkov:2004}.
This code partly utilizes the scheme originally proposed and implemented in the 
publicly available 1D code RATRAN \citep{Hogerheijde:2000}. The idea of the 
method is to solve the system of LRT equations with the Accelerated 
$\Lambda$-Iterations method, where the mean intensity $J_\nu$ is calculated 
with a Monte-Carlo approach.

We focus on pure rotational transitions and restrict ourselves to linear 
molecules with the simplest level structure (CO, HCO$^+$) in order to 
avoid the complexity caused by blends, uncertainties in collisional 
coefficients, etc. All the molecular data needed for the LRT simulation are 
taken from the Leiden Molecular
Database\footnote{{\it http://www.strw.leidenuniv.nl/\~{}moldata/}}
\citep{Shoier:2005}.

To describe and to analyze the results of the simulations, we use two 
important parameters of the transition, namely, the optical depth 
$\tau_{\nu}$ and the excitation temperature $T_{\rm ex}$. The optical depth 
is defined as
\begin{equation}
\tau_{\nu}=\int\limits_{s_0}^{s_1} \alpha_{\nu} ds,
\label{tau}
\end{equation}
where $\nu$ is the frequency within the line profile, $s_0$ and $s_1$ are 
near and far edges of the cloud along the ray, respectively, and 
$\alpha_{\nu}$ is the absorption coefficient. The optical depth 
determines how opaque the medium is and how effectively different regions 
are coupled by radiation. The excitation temperature $T_{\rm ex}$ of the transition is 
defined as
\begin{equation}
\frac{n_l}{n_u}=\frac{g_l}{g_u} \exp
\left\{ -\frac{h \nu_{ul}}{kT_{\rm ex}} \right\},
\label{texc}
\end{equation}
where $n_l$ and $n_u$ are populations of levels $l$ and $u$, and $g_l$ and 
$g_u$ are their statistical weights. The excitation temperature indicates 
how close the medium (transition) is to local thermodynamical equilibrium 
(LTE), when $T_{\rm ex} = T_{\rm kin}$. The excitation temperature and the 
optical depth can be used to constrain the LRT problem. In the optically thin 
case the temperature of the emergent radiation $T_{\rm B}$ is proportional 
to the optical depth and excitation temperature of the transition
\begin{equation}
T_{\rm B} \approx \tau T_{\rm ex}.
\end{equation}
In the optically thick case
\begin{equation}
T_{\rm B} \approx T_{\rm ex}(\tau=1),
\end{equation}
i.e., $T_{\rm B}$ is close to the excitation temperature at $\tau\approx 1$.

\subsection{Contribution Chart}

\citet{Tafalla:2006} implemented the contribution function (CF) as a 
convenient tool to analyze the relative input of a particular core region to 
the line profile. They applied it to the study of line formation and 
chemical differentiation in starless cores, using 
the frequency-integrated CF. We extend their analysis, taking into account all the
velocity channels, and investigate in detail the formation of the entire line 
profile. The contribution function is defined as
\begin{equation}
F_v(l)=e^{-\tau_v} \left\{ 1-e^{-\Delta\tau_v}\right\}S,
\label{contribition}
\end{equation}
where $\tau_v$ is the optical depth toward the element at distance $l$, 
$\Delta \tau_v$ is the optical depth of this element, and $S$ is the source 
function of the element. The index $v$ indicates that a corresponding value is 
related to a velocity offset $v$.

\clearpage
\begin{figure}[h]
\centering
\includegraphics[clip=,width=0.95\textwidth]{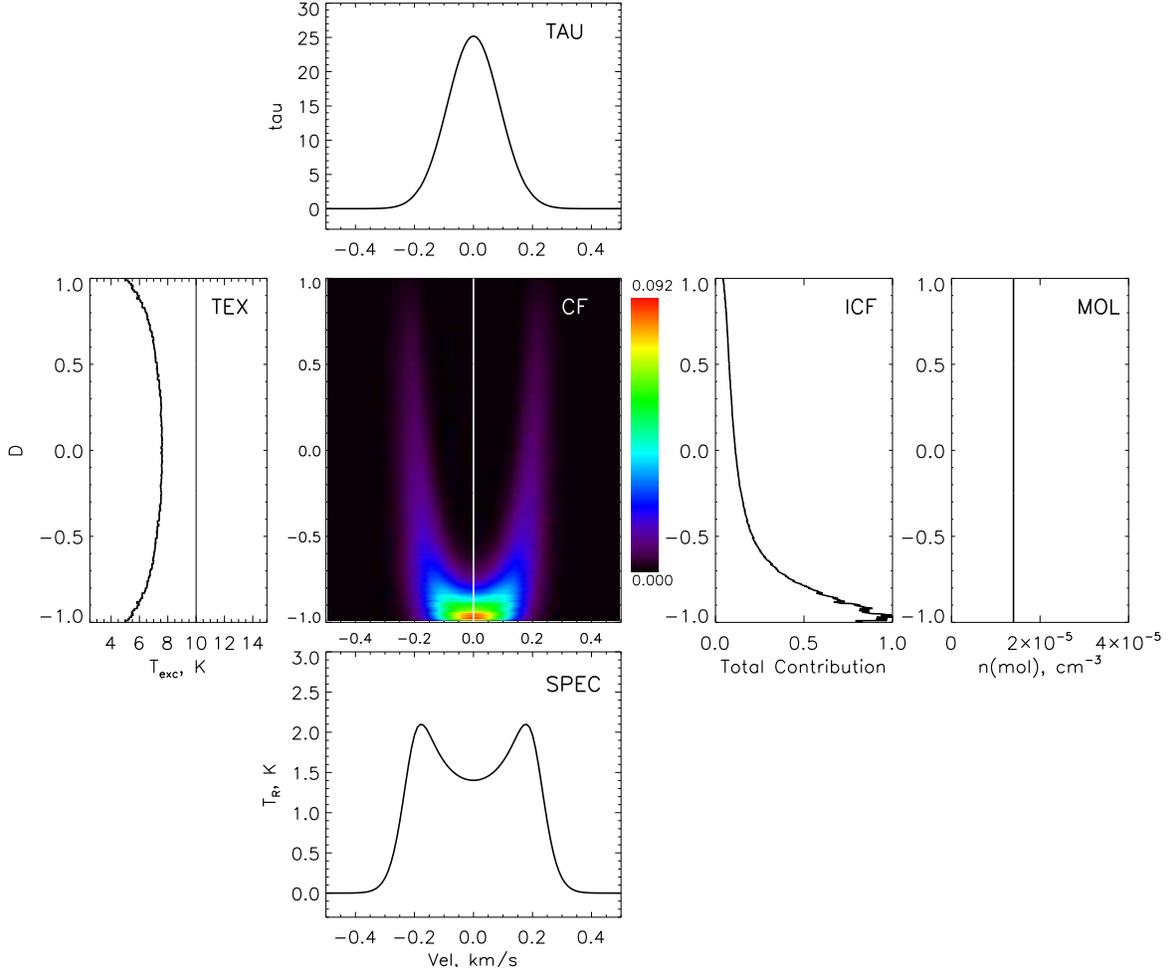}
\caption{Contribution chart for HCO$^+$(3--2) spectrum from
uniform cloud. (SPEC) Emission line profile toward the  center of the
cloud. (TAU) Spectrum of the optical depth toward the center of the
cloud. (TEX) Excitation temperature as a function of position along the
ray. The  position, $D$, is given in units of cloud radius. The value
$D=-1$ corresponds to the cloud face, while $D=1$ corresponds to the
cloud rear; (MOL) Distribution of the molecular concentration as a
function of position. (CF) Contribution function for the given position
and velocity offset. The projection of regular velocity onto the line of
sight is shown with a white line. (ICF) Contribution function integrated
over the velocity.}
\label{combine}
\end{figure}
\clearpage

We find it convenient to combine the contribution function with some other 
auxiliary plots into a common contribution chart, which is shown in Fig.~\ref{combine}. 
In this Figure we present results of HCO$^+$(3--2)
line simulations for a static uniform spherically symmetric
cloud having
$T_{\rm kin}=10$~K, $n({\rm H}_2)=10^5$~cm$^{-3}$,
$N({\rm HCO}^+)=10^{13}$~cm$^{-2}$ and $V_{\rm turb}=100$~m/s,
which are typical parameters of prestellar cores.

The emergent spectrum for the ray passing through a core center
is shown in panel~SPEC. Panel~TAU contains the 
corresponding line optical depth. Distributions of excitation temperature 
and molecular number density along the ray are shown in panels~TEX and MOL. 
The central part of the chart is occupied by a plot depicting the CF for 
the ray as a function of position and velocity (CF). 

From this chart, one readily sees that in this particular example 
radiation of the line comes almost exclusively from the nearby part of the 
cloud. Only in wings there is some weak contribution from more remote 
regions. This is emphasized in panel~ICF, where the velocity-integrated CF 
along the ray is shown. We will see later that in more realistic situations 
the picture may not be that straightforward.

\subsection{Thermalization and Critical Density}
\label{therm}

Before we start modeling the line profiles under the varios conditions,
a brief comment on the conditions in starless cores is in order. There
are two primary mechanisms  for molecules to be excited in molecular
clouds. These mechanisms are  collisions with other species (primarily
with H$_2$, He, and e$^-$) and radiative excitation. The efficiency
of the first process depends on gas  density and kinetic temperature.
The radiative excitation, in turn, depends  on the intensity of the
ambient radiation field.

\begin{figure}[h]
\centering
\includegraphics[width=0.45\textwidth]{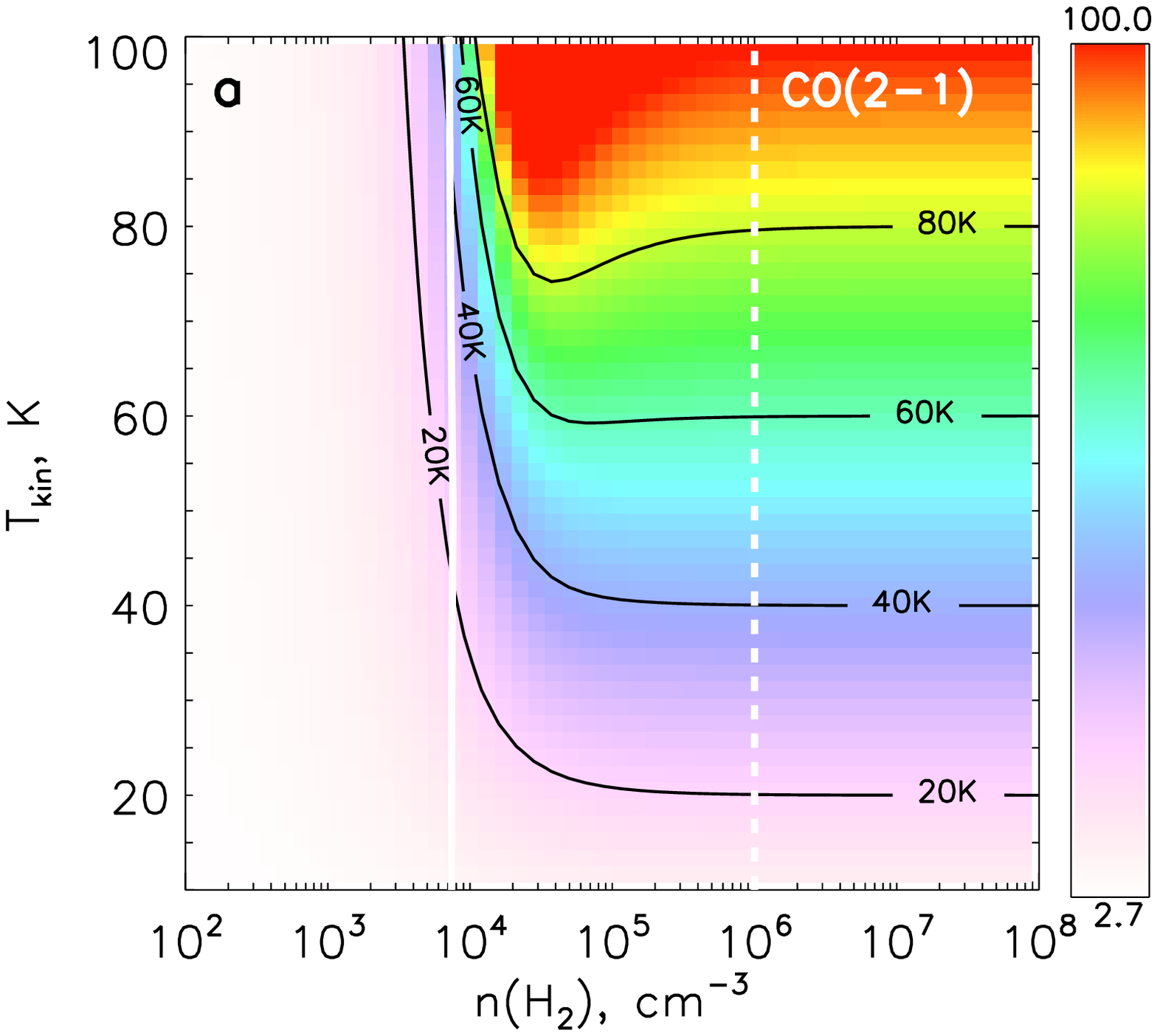}
\includegraphics[width=0.45\textwidth]{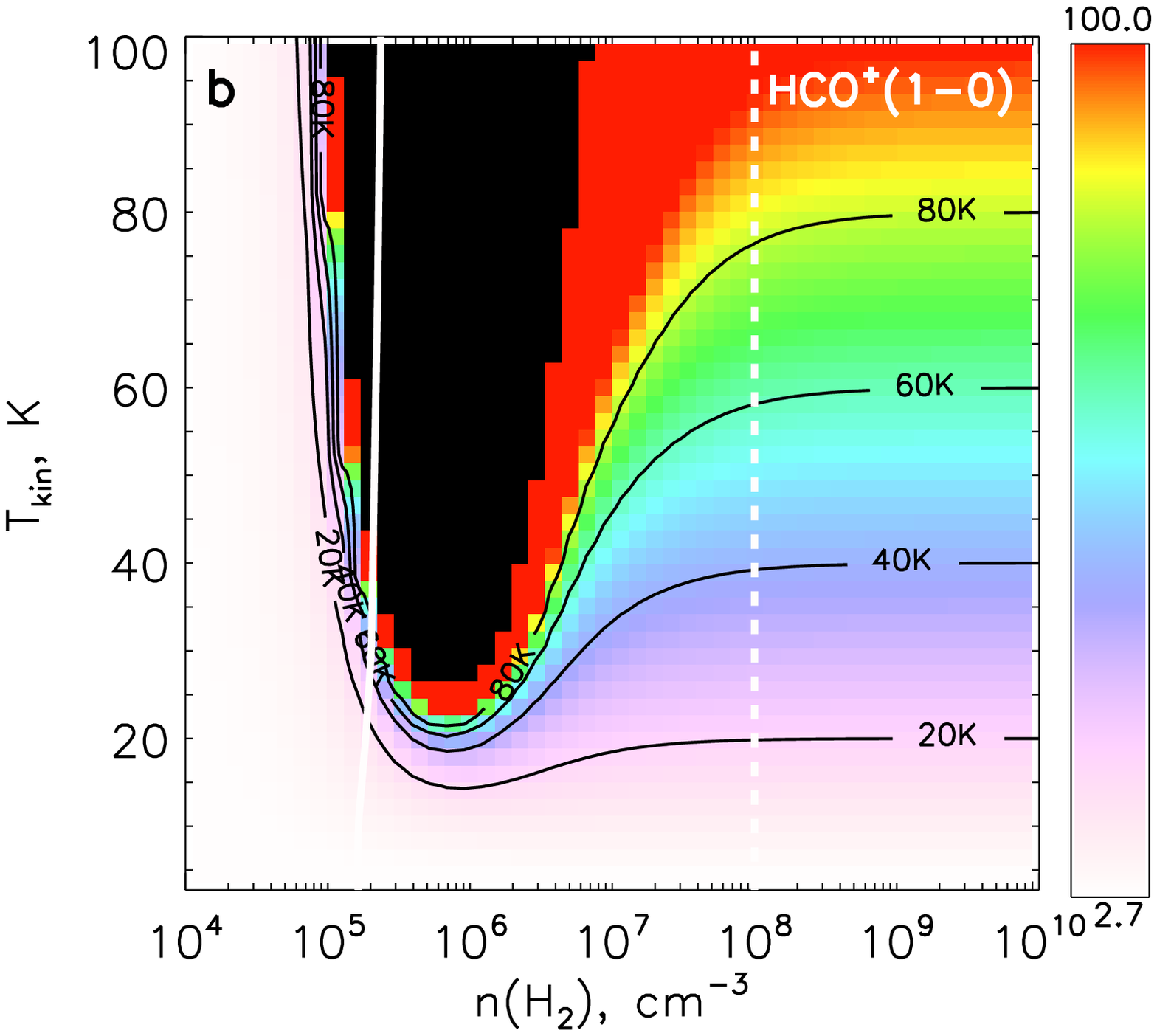} \\
\includegraphics[width=0.45\textwidth]{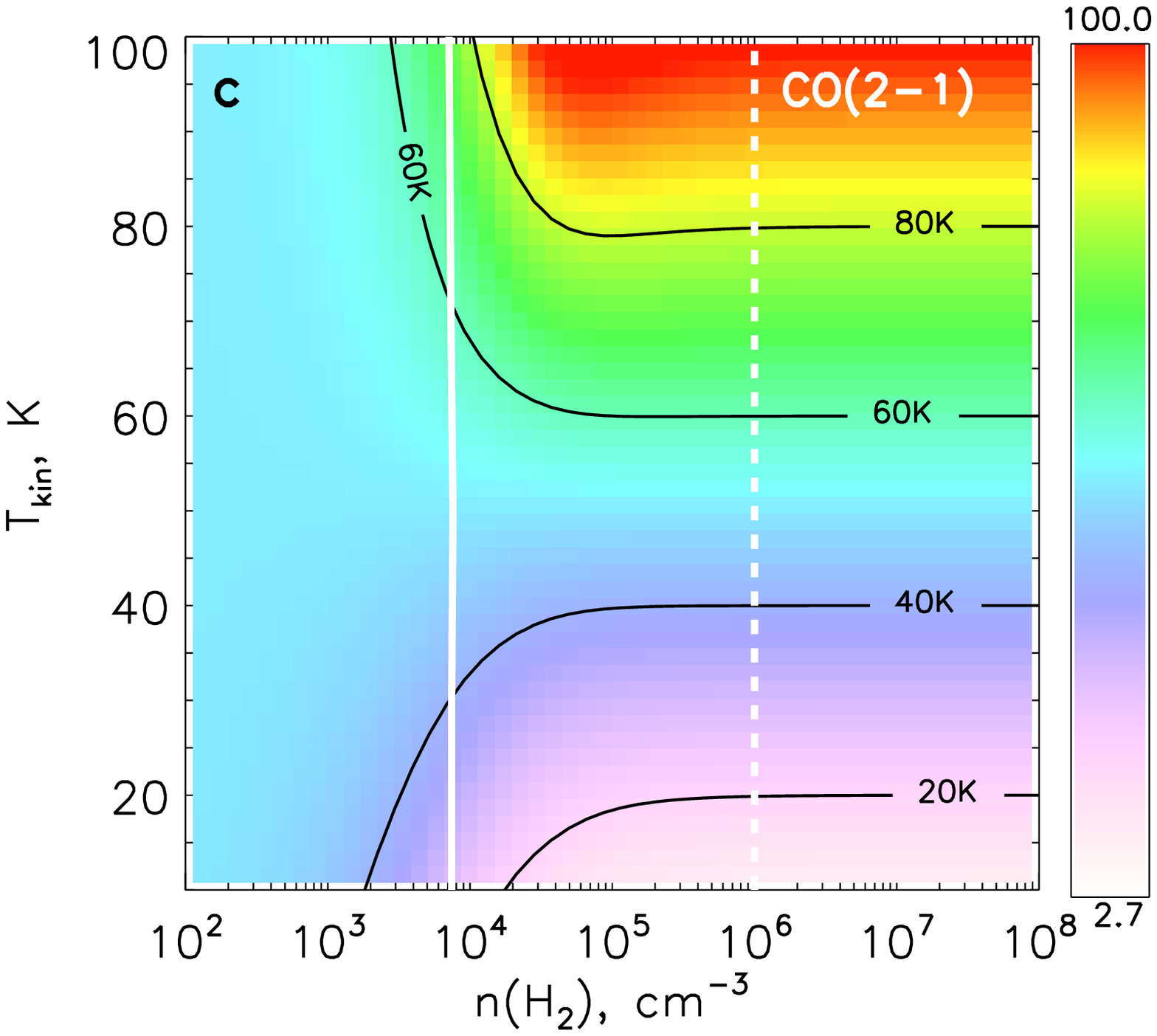}
\includegraphics[width=0.45\textwidth]{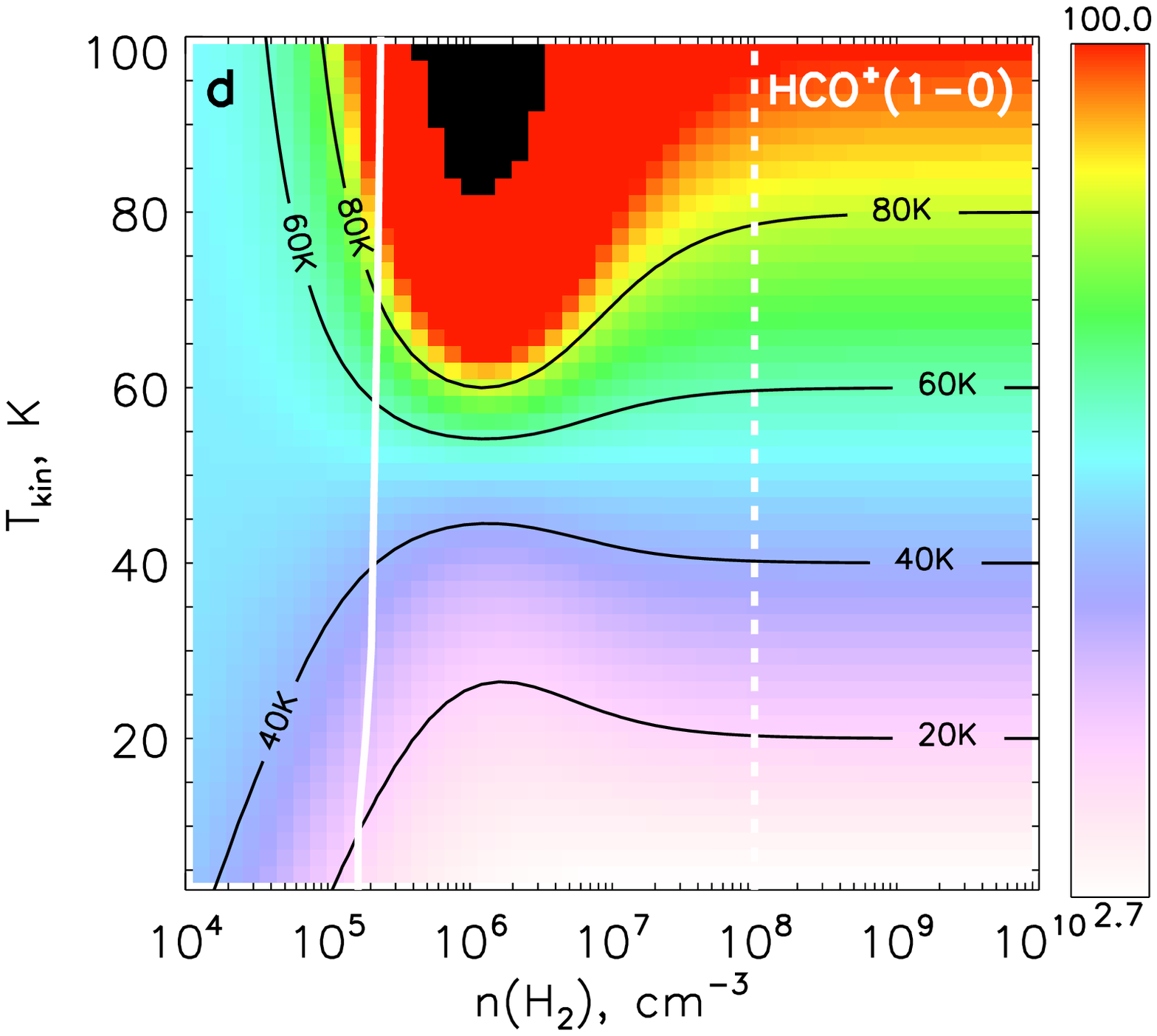}
\caption{Excitation temperature of CO(2--1) and HCO$^+$(1--0) as a function 
of hydrogen density and kinetic temperature in the optically thin limit. The 
temperature of the background radiation is $T_{\rm bg}=2.73$~K (top row) and 50~K 
(bottom row). Different colors correspond to different excitation temperatures,
as indicated on a scale located to the right of each plot.
The black area in the HCO$^+$(1--0) plots corresponds to negative 
excitation temperatures, i.e., to an inversion in level populations. 
The solid white line stands for the critical density $n_{\rm cr}$, see
Eq.~[\ref{ncr}], while the dashed white line corresponds to the thermalization
density $n_{\rm th}$.}
\label{thermal}
\end{figure}

To demonstrate the relative importance of collisional and radiative 
excitation, we consider a simple model of a uniform cloud  with
temperature $T_{\rm kin}$ and molecular hydrogen density  $n({\rm
H}_2)$, which is illuminated by isotropic blackbody radiation. The
temperature of background radiation is set to be  $T_{\rm
bg}=2.73$~K (first case) and $T_{\rm bg}=50$~K (second case),  while
$T_{\rm kin}$ is varied between 10~K and 100~K.  For our analysis, we
assume that the cloud is extremely optically thin ($\tau\ll1$), i.e. we
neglect the self-radiation of the cloud to show the effect of
radiation and collisions more clearly. We discuss the situation for two
molecules with low and large dipole moments, namely, for CO and HCO$^+$
(note, that their lines, as a rule, are not optically thin in real prestellar cores).
Excitation temperatures of CO(2--1) and HCO$^+$(1--0) as functions of
hydrogen density and kinetic temperature are presented in 
Fig.~\ref{thermal}.

\clearpage
\clearpage

It can be clearly seen that the map can be divided into three different parts. 
If the hydrogen density is lower than the so-called ``critical density'' $n_{\rm 
cr}$,
\begin{equation}
n_\mathrm{cr}=\frac{A_{ul}}{\sum_iC_{ui}},
\label{ncr}
\end{equation}
the molecular excitation temperature is determined by the radiation field. In 
Eq.~(\ref{ncr}), $A_{ul}$ is the Einstein coefficient and $C_{ui}$ are 
collisional rates from the upper level $u$ to other low-lying levels $i$. At 
densities lower than the critical density, collisions are not effective to 
excite the molecules. In this region (left part of all plots in 
Fig.~\ref{thermal}) $T_{\rm ex} \approx T_{\rm bg}$ and, thus, does not 
depend on $T_{\rm kin}$. Note that $n_{\rm cr}$ is different for different 
transitions. In our case, $n_{\rm cr}\approx 10^4$~cm$^{-3}$ for CO(2--1), and 
$n_{\rm cr}\approx 10^5$~cm$^{-3}$ for HCO$^+$(1--0). Also, $n_{\rm cr}$ is 
usually greater for upper transitions of the same molecule.

We also introduce the thermalization density $n_{\rm th}$, 
as the density, above which collisional transitions define 
level populations, so that levels are ``thermalized'', and $T_{\rm ex} 
\approx T_{\rm kin}$ (right part of all plots in Fig.~\ref{thermal}).
More precisely, we define $n_{\rm th}$ as a density at which the relative
difference between $T_{\rm ex}$ and  $T_{\rm kin}$ is 5\%.

At intermediate hydrogen densities, $n_{\rm cr}<n({\rm H}_2)<n_{\rm th}$, 
both collisional and radiative transitions effectively populate and 
depopulate levels, so that the excitation temperature is equal neither to 
$T_{\rm kin}$, nor to $T_{\rm bg}$. It can even be negative because of 
specific ratios between collisional excitation rates. The important point, 
which we would like to make here, is that these intermediate densities are 
just typical densities of starless cores. This emphasizes the necessity to 
use adequate methods for interpretation of observed molecular lines.

\subsection{Characterization of Line Profiles}

\begin{figure}[h]
\centering
\includegraphics[width=0.45\textwidth]{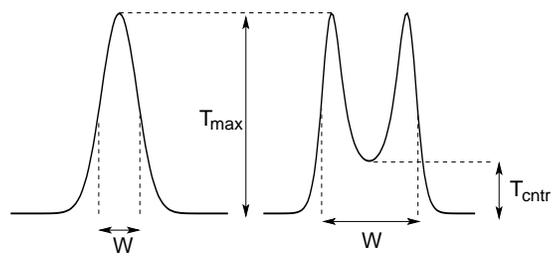}
\caption{Various parameters used to characterize line profiles.}
\label{char}
\end{figure}

In order to perform a qualitative analysis of line profiles, it is
necessary  to introduce some values to characterize them. There are
different ways to  describe numerically a given line profile with a few
parameters. In our  study we will use the peak intensity $T_{\max}$, the total
line intensity $J$ (zero moment), the mean velocity $V_{\rm cntr}$ (first
moment), the intensity at mean velocity $T_{\rm cntr}$, and the total line
width $W$ (second moment).  We put a factor $(8 \ln 2)$ in the
expression for $W$ in order to make $W$ equal to HPBW for Gaussian
profiles. In addition, we introduce the relative strength of the
self-absorption dip for  symmetric profiles, DIP. Definitions for these
quantities are given in  Table~\ref{express} and illustrated in
Fig.~\ref{char}.

\clearpage
\begin{table}[t]
\caption{Parameters of line profiles}
\label{express}
\begin{tabular}{ll}
\hline
Parameter & Definition\\
\hline
Peak intensity, $T_{\max}$ &$\max\,T_{\rm R}(V)$ \\
Total intensity, $J$ &$\int\limits_{-\infty}^{+\infty} T_{\rm R}(V) dV$ \\
Mean velocity, $V_{\rm cntr}$ &$J^{-1}\int\limits_{-\infty}^{+\infty} 
T_{\rm R}(V) VdV$ \\
Intensity at mean velocity, $T_{\rm cntr}$ &$T_{\rm R}(V_{\rm cntr})$ \\
Total line width, $W$ &$\sqrt{(8 \ln 2)\, J^{-1}\int\limits_{-\infty}^{+\infty} 
(V-V_{\rm cntr})^2 T_{\rm R}(V) dV}$ \\
Self absorption dip, DIP &$(T_{\max}-T_{\rm cntr})/T_{\max}$\\
\hline
\end{tabular}
\end{table}
\clearpage

\clearpage

\section{Molecular lines from uniform clouds}

In the previous section we showed that hydrogen density, gas temperature, 
and background radiation are all responsible for line excitation.
In the optically thin medium, the LRT problem  is in some sense local, i.e.,
the excitation temperature is completely defined by local values of $n({\rm H}_2)$, $T_{\rm 
kin}$, and by the global radiation field. However, as soon as the total column density of 
emitting molecules is large enough (i.e., lines become optically thick), the 
radiation from the entire cloud plays a role in the molecular excitation. 
Formation of line profiles in the optically thick regime is a complex 
non-local problem even in the simple case of a static uniform spherically symmetric 
cloud. 

In general, to describe the structure of such a simple cloud one specifies 
its radius $R$, temperature, $T_{\rm kin}$, micro-turbulent velocity, 
$V_{\rm turb}$, hydrogen density, $n({\rm H}_2)$, and the number density of 
a molecule, $n({\rm mol})$. However, the solution of the LRT problem depends 
on $R$ and $n({\rm mol})$ only via their product $R\times n({\rm mol}) = 
N({\rm mol})$\footnote{
It is even possible to use 
$N({\rm mol})/W$ as an independent parameter, where $W$ is the velocity
dispersion \citep[see, e.g., ][]{Lucas:1974}}. 

Therefore, spectra of uniform clouds with different radii and 
different molecular abundances but with the same molecular column density 
(other parameters being fixed) should be identical. Values of $n({\rm 
H}_2)$, $T_{\rm kin}$, and $N({\rm mol})$ are sufficient to get a unique 
solution of the LRT problem in a uniform cloud (at some velocity $v$). From 
the point of view of balance equations, $n({\rm H}_2)$ and $T_{\rm kin}$ are 
needed to calculate collision rates, while $N({\rm mol})$ determines the 
mean radiation field (in the absence of background radiation). Kinetic 
temperature and micro-turbulent velocity affect the line width.

To show how basic properties of emergent spectra depend on $n({\rm H}_2)$, 
$T_{\rm kin}$, and $N({\rm mol})$, we generate a series of models with 
different parameter sets and perform LRT simulation for each of them. Then, 
for each model we calculate the spectrum toward the center of the cloud and 
extract the main line characteristics (see Table~\ref{express}). As three 
physical parameters control the line formation in this setup, we vary two of 
them, keeping the third parameter constant. We present calculated line 
characteristics as a function of the varied parameters to illustrate their 
cooperative effects, assuming that all these parameters are independent in 
the considered ranges. Molecular abundance $X(mol)=n({\rm mol})/n({\rm 
H}_2)$ and temperature $T_{\rm kin}$ are assumed to be uniform over the 
cloud. For the analysis, we use the HCO$^+$(3--2) line which is commonly used in 
observations to probe dense parts of starless cores.
{\it Models in this Section are by no means intended to represent real
starless cores. We use them only to demonstrate the main LRT effects.}

\subsection{Effects of hydrogen density and molecular column density}

\begin{figure}[h]
\centering
\includegraphics[width=0.75\textwidth]{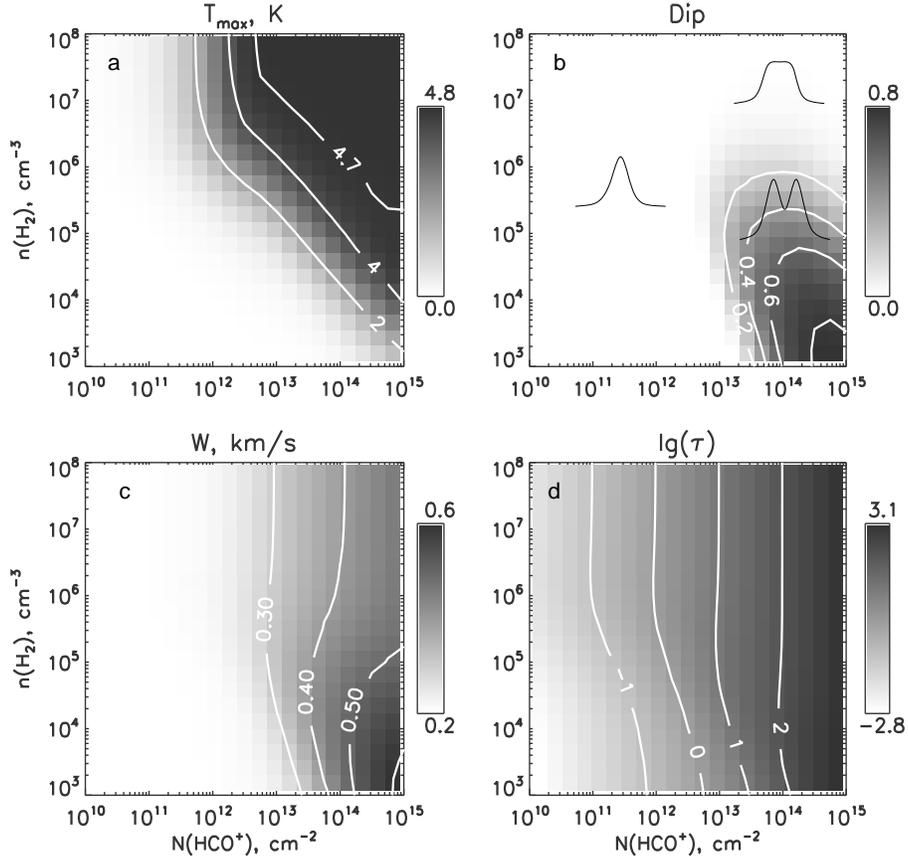}
\caption{
Greyscale maps of the line parameters for a uniform cloud.
a) Peak intensity,  $T_{\max}$, b) relative depth of the 
self-absorption dip, DIP, c) total line width, $W$, and d) logarithm of
optical depth, $\lg(\tau)$ of the emergent HCO$^+$(3--2) spectrum as a
function of $n({\rm H}_2)$ and  $N($HCO$^+)$. 
The kinetic temperature is fixed at 10~K.}
\label{column-uniform}
\end{figure}

In this subsection we consider results of $n({\rm H}_2)$ and $N($HCO$^+)$ 
variations at a fixed temperature $T_{\rm kin}=10$~K. The cloud is static 
but has a micro-turbulent velocity $V_{\rm turb}$=100~m/s, which is a 
typical value for starless cores. The basic characteristics of an emergent 
spectrum of HCO$^+$(3--2) toward the cloud center are presented in 
Fig.~\ref{column-uniform}.

\clearpage
\clearpage

When both hydrogen density and molecular column density are high 
(Fig.~\ref{column-uniform}a, top right corner), $T_{\max}$ is close to the 
LTE value ($T_{\rm R}\approx 4.8$~K for HCO$^+$(3--2)). At low $n({\rm 
H}_2)~$ and/or $N({\rm mol})$ the intensity is also low. 
However, this plot demonstrates that the common perception of a molecule 
tracing a specific density is somewhat oversimplified. The comparison of 
Fig.~\ref{column-uniform}a and Fig.~\ref{column-uniform}b shows that the 
HCO$^+$(3--2) profile can be both bright and self-absorbed even if the gas 
density is much lower than $ n_{\rm cr}$, provided the molecular column 
density is high. 

Of course, this situation may not be realistic for
HCO$^+$(3--2) in prestellar cores as at typical HCO$^+$ abundances of
about few times $10^{-9}$ \citep{ohishi} the column density of
$10^{14}$~cm$^{-2}$ and the hydrogen density of $10^{4}$~cm$^{-3}$
correspond to a linear extent of about 3~pc. But, it can be relevant for
other molecules and/or transitions, e.g. for CO(3--2). With a column
density $N({\rm CO})=10^{17}$~cm$^{-2}$ (which corresponds to an
optical depth of about 100), density $n({\rm H}_2)=10^{3}$~cm$^{-3}$
(below critical), and molecular abundance  $X({\rm CO})=10^{-4}$, the
linear extent of the cloud is about 0.3~pc, which is comparable to
typical sizes of prestellar cores.

On top right panel of Fig.~\ref{column-uniform}, the relative strength of 
the self-absorption dip is shown. As long as ${\rm DIP}=0$, the profile is 
single-peaked, otherwise it is double-peaked. The dip in HCO$^+$(3--2) line 
appears when the molecular column density is larger than $10^{13}$~cm$^{-2}$.
This corresponds to an optical depth $\approx$1, and a hydrogen 
density $n({\rm H}_2) \le n_{\rm cr}$. If the density is greater than the critical 
density, the line has a single peak even at very high HCO$^+$ column density. 
The self-absorption profile forms when a gradient in the excitation 
temperature is present. But at high density the excitation temperature is 
equal to the kinetic temperature over the whole cloud (which is assumed to 
be uniform). Thus, in such a situation we obtain a flat-top profile instead 
of a self-absorbed profile.

In the bottom panels of Fig.~\ref{column-uniform}, the total line width of the 
spectra and the optical depth are shown as functions of $n({\rm H}_2)$ and $ 
N({\rm mol})$. Both parameters are mostly determined by the molecular column 
density and only weakly depend on gas density.

We consider a broad range of HCO$^+$ column densities
(five orders of magnitude), with lower values being 
not typical for prestellar cores. Our reason is two-fold.
First, we intend to show the effects of LRT in general.
Second, the presented plots can be relevant for the isotopomers
of HCO$^+$, e.g. for H$^{13}$CO$^+$ and HC$^{18}$O$^+$
which have similar excitation parameters but smaller
abundances (defined by both isotopic ratios and isotope-dependent
chemistry).

\subsection{Effects of kinetic temperature and hydrogen density}

\begin{figure}[h]
\centering
\includegraphics[width=0.75\textwidth]{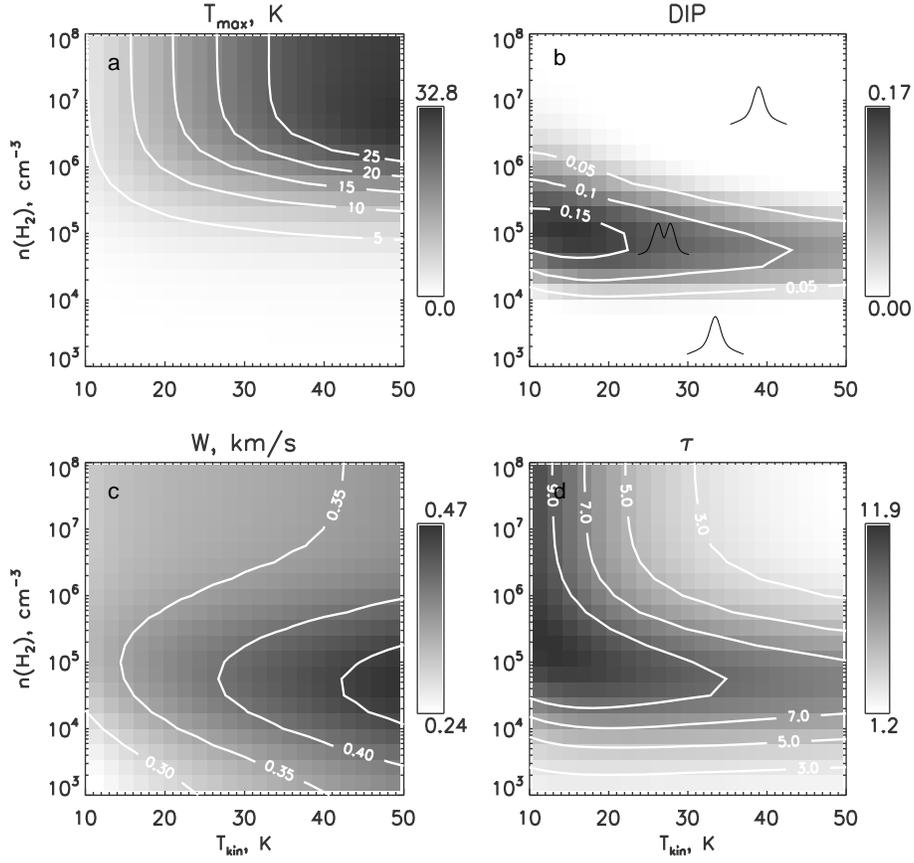}
\caption{Line parameters as functions 
of $n({\rm H}_2)$ and $T_{\rm kin}$ at 
a fixed column density $N({\rm HCO}^+)=10^{13}$~cm$^{-2}$ for a uniform 
cloud. a) Peak intensity, $T_{\max}$, b) relative depth of the 
self-absorption dip, DIP, c) total line width, $W$, and d) optical depth,
$\tau$, of the emergent HCO$^+$(3--2) spectrum are shown.}
\label{kinetic-uniform}
\end{figure}

In Fig.~\ref{kinetic-uniform} we show the combined effect of temperature and 
hydrogen density variations, keeping the molecular column density constant. 
To show the complexity of the LRT effects, we explore the case of a moderate optical
depth, which corresponds to a representative value of $N({\rm HCO}^+)=10^{13}$~cm$^{-2}$,
as follows from the previous subsection. This value is close to that inferred for the CB17 
globule in \cite{Pavlyuchenkov:2006}. To test the sensitivity of the considered 
parameters, we vary the temperature in a range, which is wider than is inferred 
for starless cores, however, all the patterns described below are seen at 
$T_{\rm kin}<20$~K.

\clearpage
\clearpage

The increase in temperature as well as the increase in hydrogen density leads to 
an increase of the line intensity. At 10~K, the self-absorption dip only 
appears at densities between $10^4$ and $10^6$~cm$^{-3}$.
At densities below $10^4$~cm$^{-3}$ collisional excitation is too slow to populate
levels of the transition, and the optical depth
is also correspondingly low. At larger densities ($>10^6$) levels higher than 3
are populated more efficiently, while lower levels are thermalized.
Because of this fact, population of lower levels decreases, which results in 
a low optical depth for densities above $10^6$~cm$^{-3}$.
The line width is sensitive to the self-absorption dip and thus
has its maximum at densities between $10^4$ and $10^6$~cm$^{-3}$.
When temperatures are higher, the density range, at which the dip appears,
gets narrower, and the dip itself 
becomes less pronounced, while the line becomes wider, and its optical depth 
decreases.

Even though we consider just one particular molecule and
particular transition, the same diagrams can be obtained for other
molecules and transitions. In essence, the diagrams for
linear molecules (for the same transitions) look similar.
The difference is that features of the diagrams are shifted along
the corresponding axes reflecting different critical densities
and different column densities per unit optical depth. Such diagrams
for more complicated models can be used to illustrate the main 
non-LTE effects of the line formation process and to derive parameters
of prestellar cores from the observed spectra.

\section{Molecular line formation in starless cores}

Even in uniform clouds and in the absence of regular velocity field,
line shapes can be quite different from simple gaussians, having
self-absorption dips or flat tops due to thermalization  (see above). Of
course, when regular velocity fields and chemical differentiation are
taken into account, line profiles become even more diverse, which
further complicates their analysis. In this Section we consider the
formation of the HCO$^+$(1--0) line in the presence of chemical
stratification and regular motions, using a simple
``chemo-dynamical''\, model, where the chemical evolution is calculated
along with the dynamical evolution of the core.

\subsection{Model of starless core}

We assume that a starless core is spherically symmetric with the
density distribution given by $n(r)=n_0/(1+(r/r_0)^p)$, where $p=2.4$,
$r_0=2800$~AU, $n_0=1.35\times10^6$~cm$^{-3}$. Parameters are chosen to represent
the L1544 core \citep{Tafalla:2004}. To calculate
the chemical and kinematic structure of a starless core, we
use the model described in \cite{Pavlyuchenkov:2006}.

We consider models of a static core, collapsing core, and rotating core.
To quantify collapse and rotation velocities, we assume that the model
core forms over a time $t_0=0.8$ Myr as a result of contraction from an
uniform initial configuration with density $n_{\rm
ini}=5\cdot10^3$~cm$^{-3}$ and rotation velocity at the core edge
$V_0=100$ m s$^{-1}$. The micro-turbulent velocity is 100~m s$^{-1}$ in
the static model and 50~m s$^{-1}$ in the non-static models. 
Molecular abundances are calculated with the same chemical model as described in
\cite{Pavlyuchenkov:2006}. Briefly, it is a time-dependent chemical
model which includes gas-phase reactions as well as the freezing-out of
molecules onto dust  grains and their desorption back to the gas-phase.
Surface reactions are not taken  into account. Species stick to dust
grains with the probability of 0.3 for all components except for H$_2$
and He, for which zero sticking is assumed. We calculate the sticking
probability for atomic hydrogen according to \cite{hmk}. Three
desorption mechanisms are taken into account, which are thermal
desorption, photodesorption, and cosmic ray induced desorption. It is
assumed that a cosmic ray particle impulsively heats a dust grain to a
peak temperature $T_{\rm crp}$, which is close to 70~K for cold
$0.1\mu$m silicate grain. Data from \cite{leger} are used to
approximate the dependence of $T_{\rm crp}$ on the initial dust
temperature \citep{swh2004}. Gas-phase reactions are taken from the
UMIST\,95 ratefile \citep{umist95}. We consider the evolution of species
containing H, He, C, N, O, Mg, Na,  Fe, S, and Si atoms.

Kinetic temperature is assumed to be 9~K. The core is subject to an
external UV field with the intensity, $G$, measured in units of the mean
interstellar field \citep{Draine}, and to cosmic rays, which ionize
molecular hydrogen with the rate $\zeta$. The assumed distance to the
core is 140~pc as believed to be relevant for cores in Taurus
cloud complex (which in fact maybe closer; see note in \cite{taurus}).
Using these parameters we perform LRT simulations and produce ``ideal''\, 
(non-convolved) spectral maps and contribution charts for the
HCO$^+$(1--0) line.

\begin{figure}[h]
\centering
\includegraphics[width=0.5\textwidth]{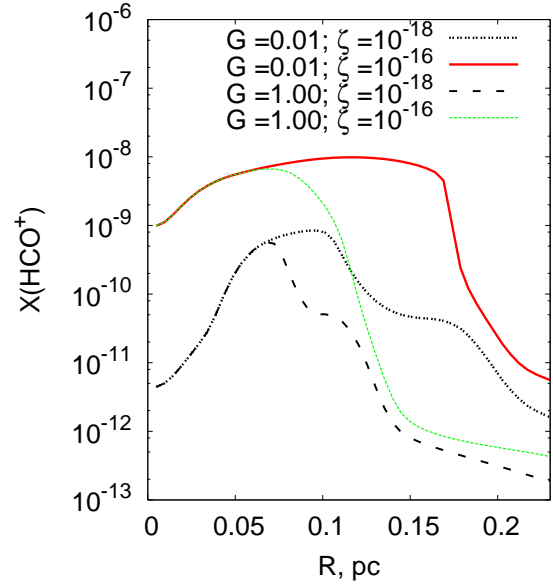}
\caption{Distribution of the HCO$^+$ relative abundances in the
prestellar core model for the different parameters of UV radiation and CR
flux.}
\label{hco+}
\end{figure}

To illustrate the importance of the cosmic ray (CR) ionization rate and
ultraviolet (UV) radiation intensity for the chemical structure of the
core in Fig.~\ref{hco+} we present abundance profiles for HCO$^+$
molecules calculated with $G=0.01$ and 1.0 and $\zeta=10^{-18}$ s$^{-1}$
and $10^{-16}$ s$^{-1}$ at $t=0.8$~Myr.

\clearpage
\clearpage

The relative abundance of HCO$^+$ strongly depends on
both parameters and varies up to a few orders of magnitude. The strong
UV radiation destroys molecules in the envelope, while enhanced CR flux
prevents molecules from sticking to dust grains (via CR-induced desorption)
and enhances HCO$^+$ abundance over the whole cloud. 

\subsection{Static core}

\begin{figure}[h]
\centering
\includegraphics[width=0.8\textwidth]{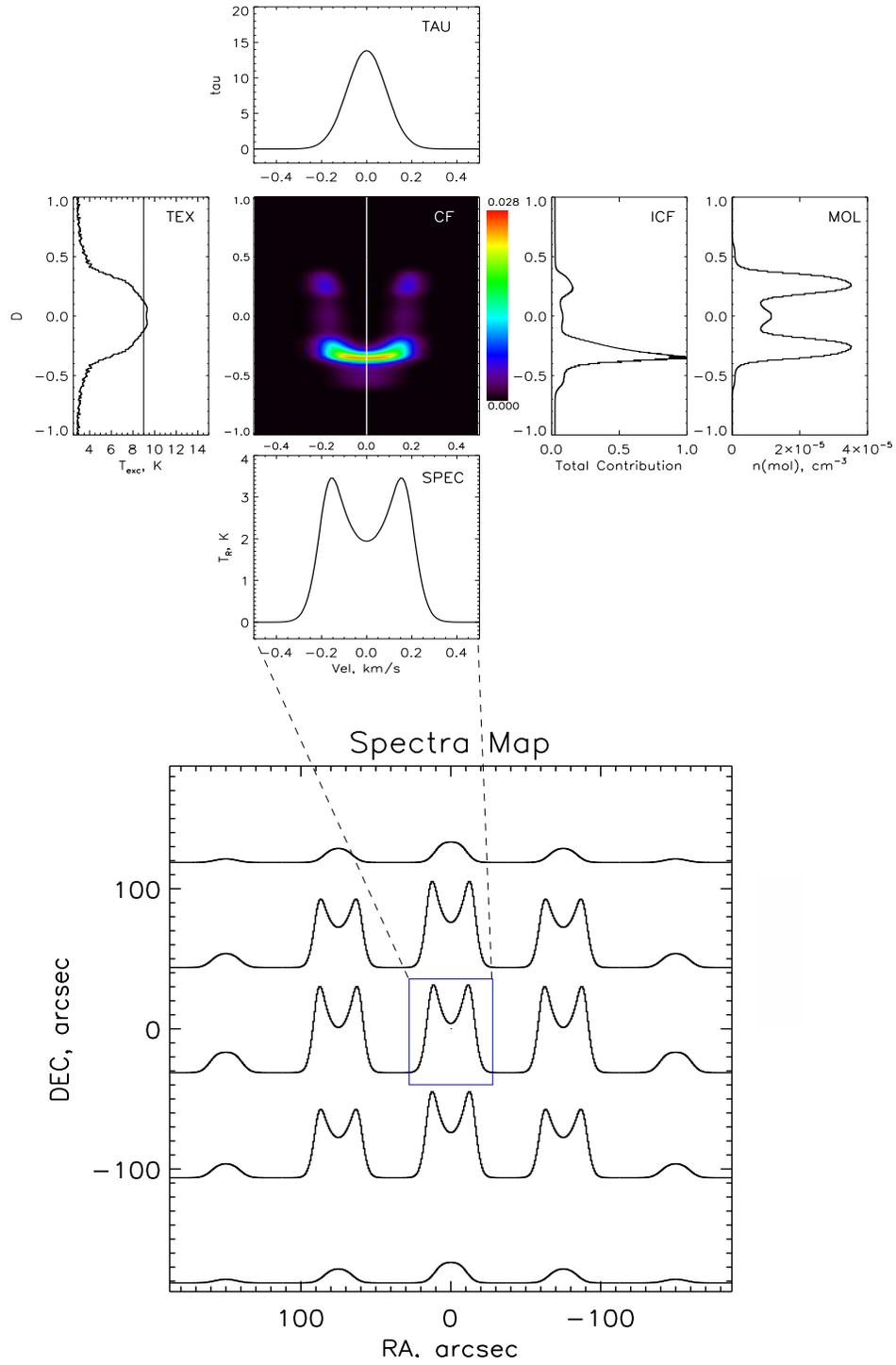}
\caption{Top: Contribution chart for the central spectrum of
the non-uniform static core. Bottom: Spectral map
of HCO$^+$(1--0) for the same core.}
\label{explain1}
\end{figure}

We first assume that the core is static, i.e., both infall and
rotation velocities are zero. The calculated spectral
map and the contribution chart for the central
spectrum are shown in Fig.~\ref{explain1} for the case when $G=0.1$ and $\zeta=10^{-18}$~s$^{-1}$.

\clearpage
\clearpage

The contribution chart demonstrates all the aspects of HCO$^+$(1--0) line formation.
In the inner part of
the core the excitation temperature $T_{\rm ex}$ is nearly equal to the kinetic
temperature while in the envelope it decreases with radius (TEX). This is
mainly because of the lower density in the envelope, so that collisions
are not effective enough to excite the level populations. The molecular
density distribution is not uniform but has a maximum at $R/R_0=0.2$ (MOL).
In the inner part the core molecules are depleted onto the dust grains. The optical depth in
the line center is about 20, i.e., the line is optically thick (TAU).
From the chart, it can be seen that emission in the line center
comes from the outer part of the core where the excitation temperature
is relatively low. The emission in blue and red peaks comes mainly from inside,
where the excitation temperature is higher. This results in the
double peaked profile. At the same time, the very center of the core does
not contribute to the HCO$^+$(1--0) profile even in line wings
because of molecular depletion.

\subsubsection{Uncertainties of the chemical structure}

A non-uniform chemical structure of starless cores is of crucial
importance for the interpretation of observed spectral maps. A particular
starless core may look very different in lines of various
molecules just because of the different molecular abundance
distributions \citep{Tafalla:2002}. In general, current models of starless cores, which
include chemical and micro-physical processes (e.g., depletion and desorption
of molecules on dust grains), allow us to understand {\it qualitatively}
distributions of simple molecules like CO, CS, HCO$^+$, and
N$_2$H$^+$ \citep[see, e.g.,][]{Bergin:2007,Aikawa:2008}.
However, it is still difficult to use such models for the {\it quantitative} analysis, mainly because
of two reasons. First, rates
of many chemical and physical processes, involved in determining
the abundance gradients, are still very uncertain. 
Here we refer to the papers by \citet{Vasyunin:2004} and 
\citet{Wakelam:2006} where the uncertainties of chemical reaction rates are studied.
Second, external physical conditions such as intensity of the CR and UV radiation
fields, are often not well constrained, but they strongly affect the
core chemical structure \citep[see, e.g.,][]{Farquhar:1994, Pavlyuchenkov:2006}. 

\begin{figure}[h]
\centering
\includegraphics[width=0.75\textwidth]{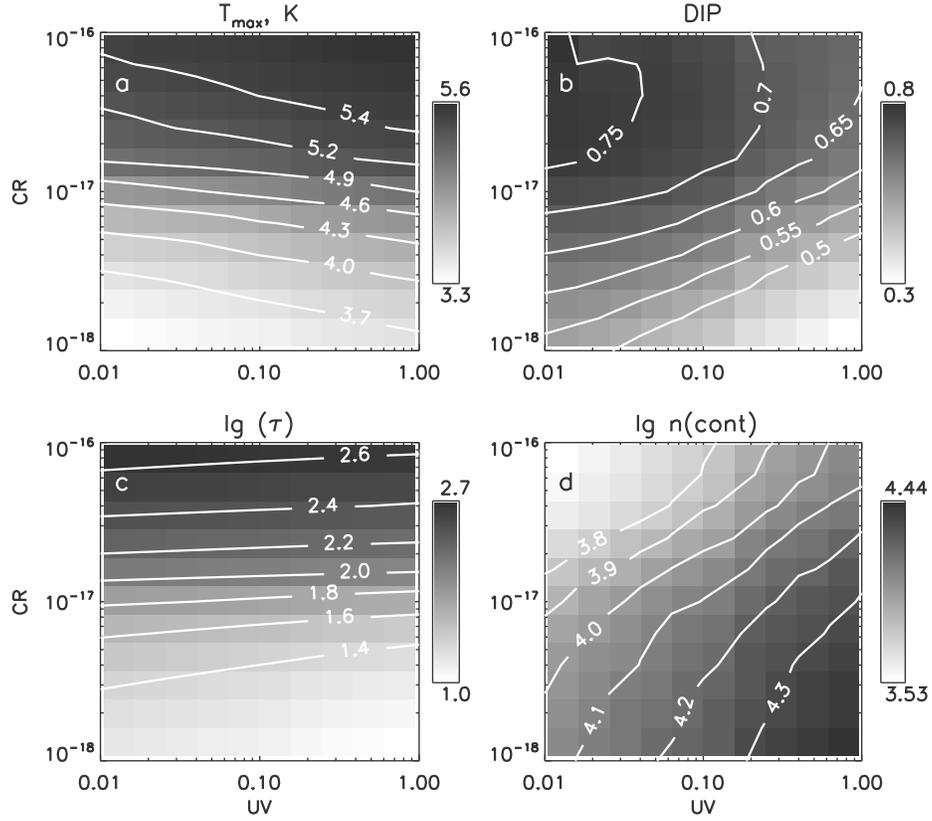}
\caption{Results of the variation of the UV and CR radiation fields for a non-uniform model
of a prestellar core. a) Peak intensity, $T_{\max}$, b) relative
strength of the self-absorption dip, DIP, c) the logarithm of optical
depth,  lg~$(\tau)$, and d) the logarithm of contributing density
$\lg n$(cont), (see text) of the line formation for HCO$^+$(1--0) are
shown as functions of $G$ and $\zeta$.}
\label{core-chem}
\end{figure}

Here we want to illustrate the influence of the $G$ and $\zeta$ values on the emergent spectra.
We consider a range of ``L1544-like'' models, but varying $G$ from 0.01 to 1 and
$\zeta$ from $10^{-18}$~s$^{-1}$ to $10^{-16}$~s$^{-1}$.
The resulting difference in abundances affects both the  intensity and the relative
strength of the self-absorption dip (Fig.~\ref{core-chem}).

\clearpage
\clearpage

In the bottom right panel we show the density of the line formation, 
or ``contributing density'', which we define as the density where the integrated
contribution function has a maximum (Fig.~\ref{explain1}-ICF). This
density approximately indicates which densities are actually ``traced''
by the considered  transition. The contributing density depends on the 
UV and CR values and can be even smaller than the critical density  (cf.
Fig.~\ref{thermal}). In the adopted parameter space the optical depth
of the HCO$^+$(1--0) transition varies from 10 to 1000.

From the point of view of the line intensity and its optical depth, $\zeta$
seems to be more important than $G$. The reason is chemistry.
In the considered parameter space,
$T_{\max}$ and $\tau$ almost do not depend on the UV field intensity,
because, when $G$ increases, the maximum of the absolute HCO$^+$ abundance
just moves deeper into core, without changing the column density
much. On the other hand, an increase in $\zeta$
affects the entire core, enhancing the gas-phase molecule abundance and
the column density.

As can be seen from Fig.~\ref{column-uniform}d, the optical depth depends strongly on
the column density and is almost independent on gas density (i.e., excitation
conditions). This is reflected in Fig.~\ref{core-chem}c. Peak intensity
and the depth of the self-absorption dip depend both on molecular column density and
H$_2$ volume density (Fig.~\ref{column-uniform}a,b). Thus, the corresponding plots in
Fig.~\ref{core-chem}a,b have a more ``diagonal''\  appearance. The dip gets deeper as the UV
intensity decreases, because at lower $G$ the maximum HCO$^+$ abundance occurs
closer to the core edge, i.e., at smaller densities. In other words, when $G$ is low,
HCO$^+$(1--0) traces lower densities (Fig.~\ref{core-chem}d).

It must be noted that the description used in this subsection is somewhat simplified
in the sense that it does not take into account the temperature gradient which may appear
due to enhanced UV irradiation and/or CR flux. Even though these small temperature
variations are of minor importance for chemistry, they do affect the excitation conditions
and, thus, further complicate the situation.

\subsection{Collapsing core}

\begin{figure}[h]
\centering
\includegraphics[width=0.8\textwidth]{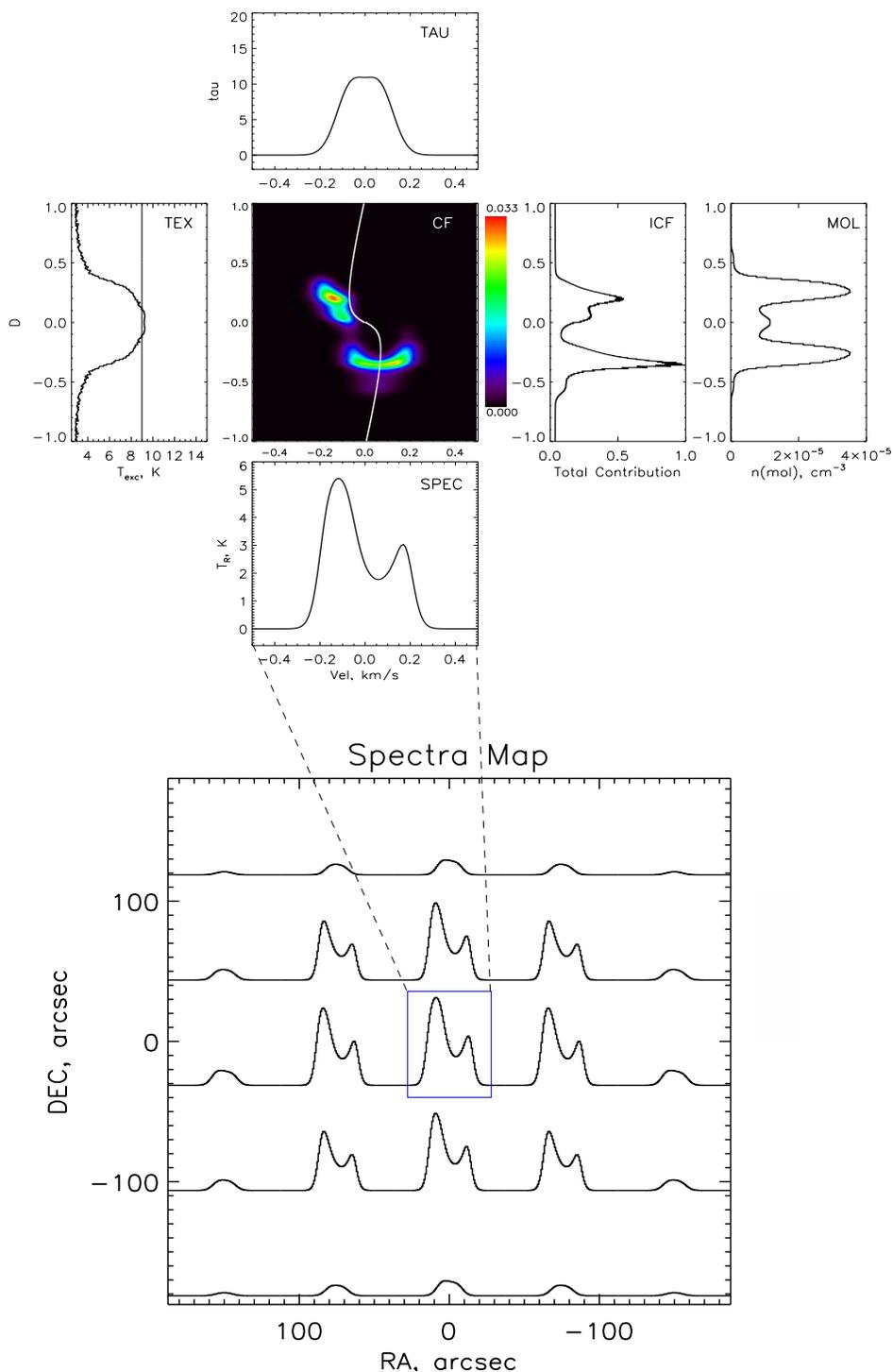}
\caption{
Contribution chart for HCO$^+$(1--0) emission from the collapsing
core with $G=0.1$ and $\zeta=10^{-18}$~s$^{-1}$. Because of strong
depletion, the line peaks mainly form in the outer parts of the core,
but there is still some contribution of the core center in the blue
peak. Projection of regular velocity onto the line of sight is shown
with a white line.}
\label{explain2}
\end{figure}

The next step is to abandon the assumption of zero regular velocity in the core.
First, we assume that the core is contracting, with a maximum
infall velocity of about 50 m/s. The corresponding spectral map is shown
in Fig.~\ref{explain2} for the case with $G=0.1$ and $\zeta=10^{-18}$~s$^{-1}$.
 
\clearpage
\clearpage

In contrast to the static model, line profiles show the classical
infall asymmetry, with the blue peak dominating over the red peak. The
map itself is symmetric relative to the center. The combination of factors behind the line profile asymmetry is well
seen on the chart. The contribution chart for
the central spectrum (Fig.~\ref{explain2}) has a complex appearance, which,
in general, follows the projected velocity (white curve in Fig.~\ref{explain2}).
There is a common perception that optically thick lines trace
outer parts of the source. However, as we see from
the chart in Fig.~\ref{explain2}, in general, the line profile includes information
about different parts of the cloud. The integrated contribution function (Fig.~\ref{explain2}-ICF) shows that
both near and far parts of the core participate in the formation of the
central spectrum. As in the static case (Fig.~\ref{explain1}), the emission around
the dip (at zero velocity) and in the red peak comes from the envelope part closest to the
observer. However, this region contributes some emission in the blue peak as well.
The most of the emission in the blue peak comes from the
rear part of the model core (a hallmark of the infall asymmetry) and also
from its center. The gap between the near and rear parts of the
contribution function (Fig.~\ref{explain2}-CF) at negative velocities
is caused by depletion of HCO$^+$ in the core center.

Self-absorption spectra toward centers of some prestellar cores
\citep[e.g.][]{Lee:1999,Lee:2001} are dominated by red peaks,
as if the corresponding core is expanding.
The asymmetry of the line profile in a particular core can also be caused
by an outflow powered by an unseen central object (protostar).
In addition, spectral maps of some cores
\citep[e.g.][]{Lada:2003} indicate that their kinematics cannot
be described as pure contraction, expansion and/or rotation.
\citet{Keto:2006} suggested that the velocity field in these prestellar
cores can be reproduced by a model of an oscillating pressure-bounded,
thermally-supported object.
 
If we, for the purpose of discussion, consider a
model of an expanding core by changing the velocity sign, we end up
with a spectral map and contribution chart for the central
spectrum, which are just mirror reflections of the same plots for the
model of contracting core but with the opposite asymmetry.  
In general, a detailed dynamical model is needed to quantify such motions.

\subsection{Rotating core}

\begin{figure}[h]
\centering
\includegraphics[width=0.8\textwidth]{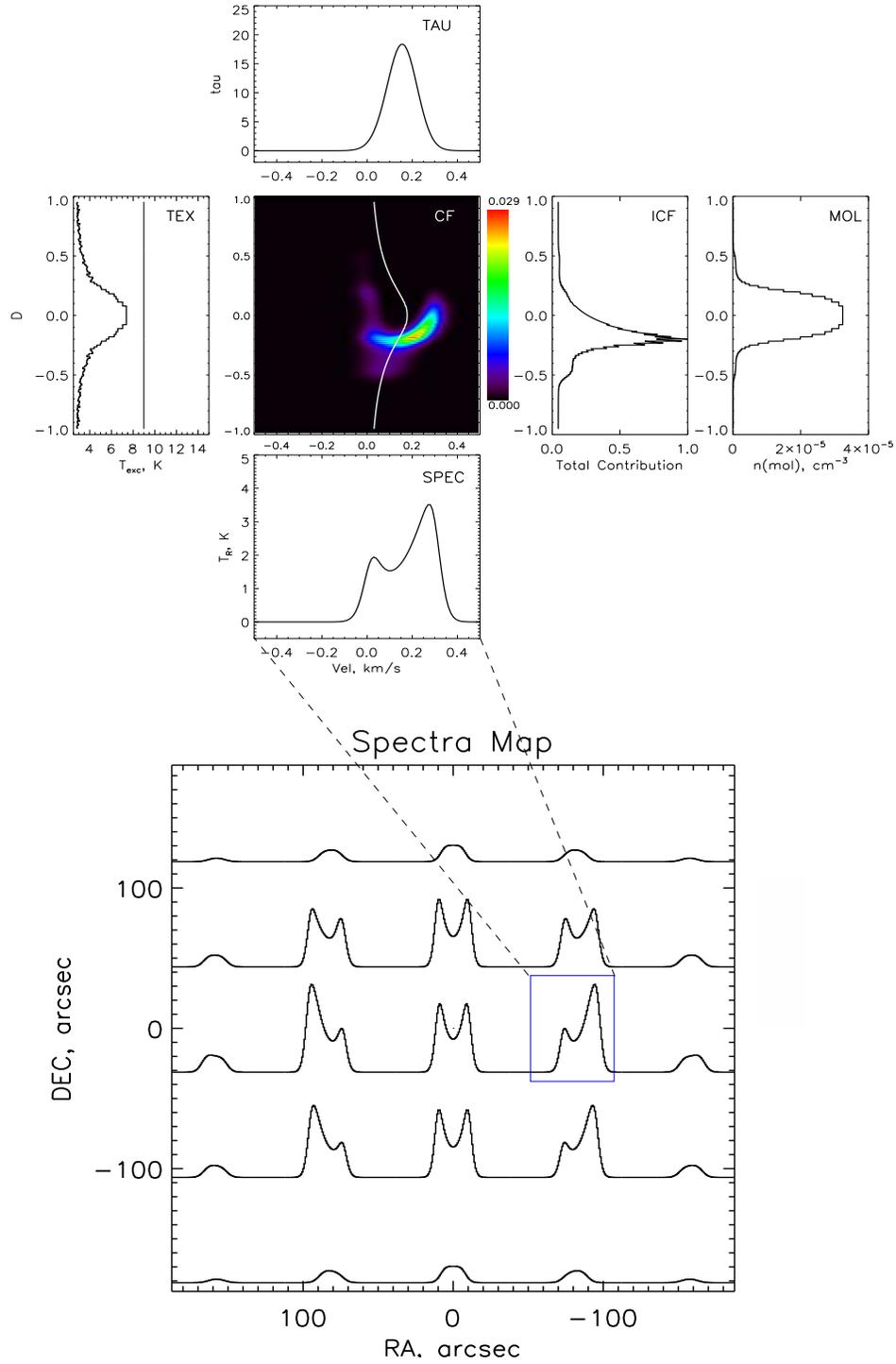}
\caption{Contribution chart for HCO$^+$(1--0) emission from the rotating core. Projection of
the rotation velocity onto the line of sight is shown with a white line.}
\label{explain3}
\end{figure}

Now, we assume that the core does not collapse, but rotate with the
axis being perpendicular to the line of sight. The
corresponding spectral map and the contribution chart for HCO$^+$(1--0) are
presented in Fig.~\ref{explain3}. The line profiles are asymmetric everywhere
in the map except for the projection of the rotation axis. 
To the left of this axis, profiles are blue-dominated and blue-shifted along the velocity,
while to its right they are red-dominated and red-shifted. The map is symmetric
with respect to the projection of the rotation axis. 
As in the case of the contracting core, the appearance of the contribution
function follows the projected velocity. 

\clearpage
\clearpage

Comparing lines from contracting and rotating models, we see that they 
produce very similar profiles in some off-central positions. Therefore,
in order to distinguish between the models it is necessary to analyze
the complete spectral map, not just the central or single off-central
spectrum. Moreover, in order to extract the information about kinematics
of the cloud it is of a great importance to analyze simultaneously
spectral maps for different transitions (optically thin and optically
thick) and different molecules.

\subsection{Collapsing and Rotating core}
\begin{figure}[h]
\centering
\includegraphics[width=0.8\textwidth]{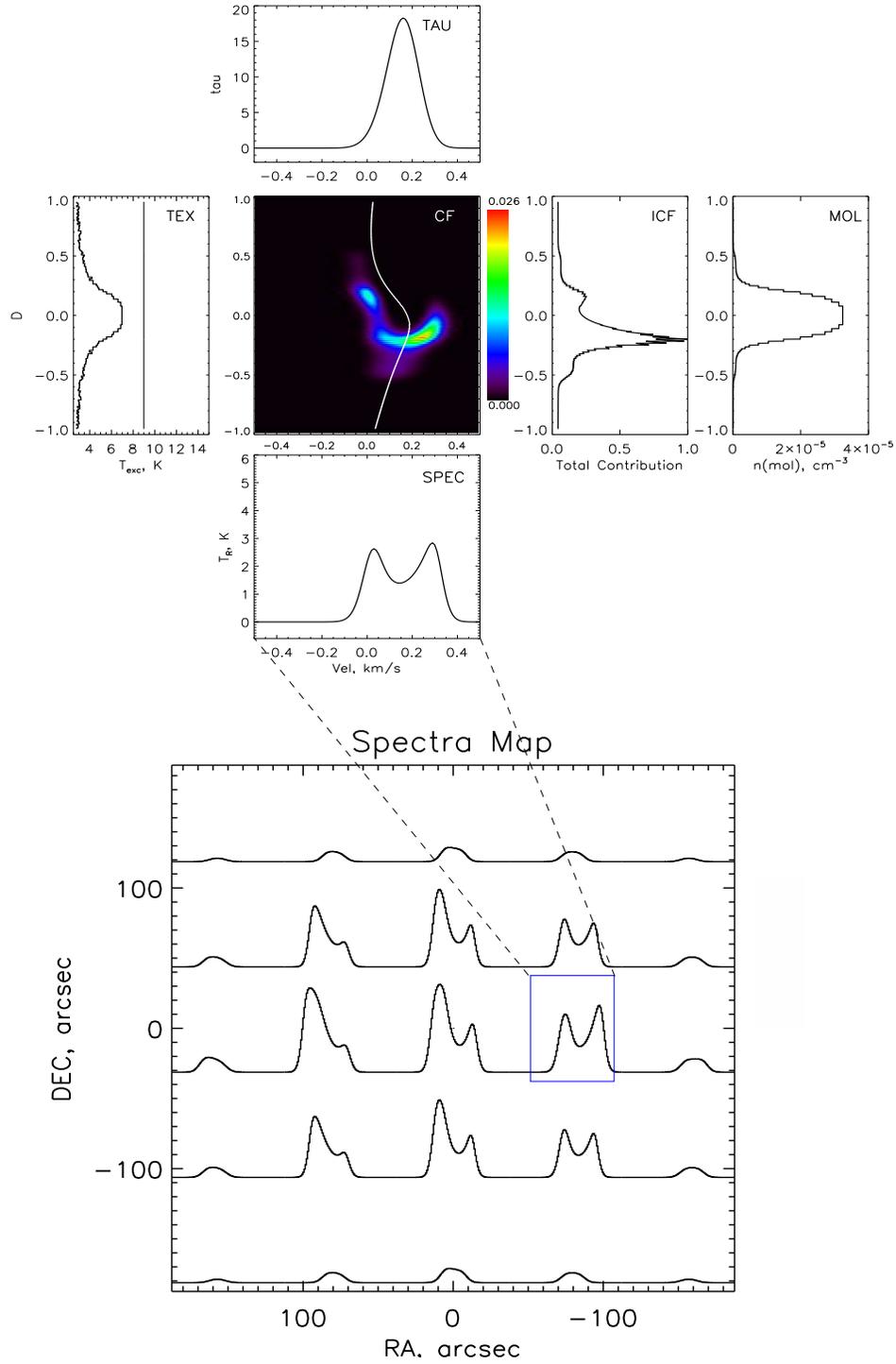}
\caption{ 
Contribution chart for the HCO$^+$(1--0) emission from a
collapsing and rotating core. The projection of the regular velocity
onto the line of sight is shown with a white line.}
\label{explain4}
\end{figure}

Finally, let us assume the more realistic case, i.e. that the core is collapsing and rotating at
the same time. The corresponding spectral line map and the contribution chart
for HCO$^+$(1--0) and $i=90^\circ$ are presented in Fig.~\ref{explain4}. The spectra map
is no longer symmetric, neither with respect to the center (as in the
case of a purely contracting core), nor with respect to the projection of the
rotation axis (as in the case of pure rotation).  On the left side
of the map, the effects of infall and rotation are coherent and profiles
are strongly blue-dominated. On the right side of the map, infall
and rotation tend to produce opposide line asymmetries
(see Fig.~\ref{explain2}--\ref{explain3}) and their net result is the
formation of nearly symmetric line profiles. To the left of the rotation
axis, profiles are blue-shifted, while to its right
they are red-shifted, as in the case of pure rotation.

\clearpage
\clearpage

Combination of collapse and rotation allows to reproduce the
spectral features observed in a number of molecular cores. In
particular, \cite{Pavlyuchenkov:2006} used a similar model for the
CB17 prestellar core and derived the infall and rotation
velocity by fitting the maps of optically thin and optically thick lines
simulteneously. The presence of both rotation and infall is also seen in
the L1544 core \citep{Ohashi:1999, Williams:2006}. To distinguish
between rotation and infall it is also useful to use the first-moment
maps, \citep[e.g.][]{walker1994,Chen:2007}.

\section{Discussion}

Starless cores arguably represent the simplest configuration among all
stages of the formation of a low-mass star. Despite this simplicity, the
interpretation of observed spectra is not straightforward, especially,
in terms of kinematics, because regular motions in these objects have
velocities which are comparable to or even less than the sound speed.

Above, we have given some examples of molecular line formation in static
and dynamic cores, demonstrating how line profiles are influenced by gas
density and temperature, molecular abundance, presence of regular
motions, and external parameters which drive the chemical evolution.
Contribution charts show clearly that any optically thick line,
generally speaking, does not represent some specific part of the core
but rather combines contributions from different regions. Incidentally,
this is what makes it possible to use optically thick lines as a
diagnostic for core-wide kinematics. On the other hand, while the
contribution function in general follows the regular velocity profile,
emission at a given velocity is not unambiguously related to a region
which moves with this velocity, bearing information from other regions
as well because of finite microturbulent line broadening.

There are some other potential complications which are not accounted for
in the present paper. In particular, in the models presented above, the
gas temperature is assumed to be uniform over the core. In reality, the
thermal structure of starless cores can be more complex. Following the
theoretical considerations, the gas temperature is a result of heating
by cosmic rays, photoelectric heating, collisional exchange with dust,
and cooling by atomic and molecular lines. The dust temperature, in
turn, is mostly controlled by external radiation and
expected to rise from inside ($\sim5-7$~K) to outside ($\sim15$~K)
\citep[see e.g.][]{zucconi, Evans:2001, Young:2004}. Cooling by
molecular and atomic lines depends on the abundance of key species, like
CO, C and C$^+$ \citep{goldsmith}. Thus, in order to estimate 
the gas temperature distribution one needs to model the chemical structure,
line radiative transfer, and the energy exchange between gas and dust.
\cite{Keto:2008} showed that the thermal structure of starless cores may
differ depending on their central density. For cores with high central
density ($n({\rm H}_2)>10^5$ cm$^{-3}$), the temperature rises from
inside to outside, while for less dense cores the temperature
distribution is more uniform, which seems to be supported by
observations of several prestellar cores. As shown by
\cite{Pavlyuchenkov:2007} over-simplified thermal models may lead to
wrong interpretations of line data. Thus, the thermal structure of the
starless cores should be considered in the LRT simulations. In
fact, the sensitivity of CO lines to the gas temperature in the
envelope makes it possible to use these lines to constrain the
UV part of the radiation field, while the dust RT modeling
allows to estimate the overall level of the interstellar radiation
field, \citep[see][]{Evans:2005}.

From the observational side, there is a convolution problem which
is related to the finite resolution of radio-telescopes. If the
telescope beam is comparable to the angular size of an object then
different regions contribute to the emergent line profile which
makes derivation of source parameters even more difficult and often
ambiguous.

There are also some uncertainties which are related to the LRT problem
itself. One of them is the commonly used assumption of the full
frequency redistribution (FFR) over the line profile which dramatically
simplifies equations of the LRT (e.g., the source function does not
depend on frequency). However, this assumption may not be valid in low
density regions where collisions are not frequent enough to redistribute
the absorbed energy. As a result, the line formation may proceed in a
way between two extreme modes, namely, between FFR and coherent
scattering. In the latter case, lines would not have any
self-absorption dips (but shapes can still be complex because of the
regular velocity).

Another commonly used approach is to simplify the non-regular
(turbulent) velocity field in cores by introducing the micro-turbulence
velocity (Eq.~[\ref{vturb}]). Such representation of the real (unknown)
velocity field by the combination of the regular and microturbulent
velocities also may lead to misinterpretations.  For instance, line
asymmetry can be considered as an evidence of infall (outflow), being
actually a reflection of a complex kinematic structure due to turbulence
or oscillations. Moreover, even if we are not interested in line shapes,
the microturbulent approach can fail in reproducing line intensities. As
shown by, e.g., \cite{Hegmann:2000}, the turbulent structure of prestellar
cores can be rather represented by meso-turbulence which again requires
the modification of the radiative transfer equation.

\section{Conclusions}

In this paper we analyze molecular line formation at
conditions typical of starless cores. In particular, we show the effect
of density, molecular column density, and temperature on line
parameters for a sample of uniform clouds. We also consider non-uniform
models and show the effect of the chemical differentiation, collapse and
rotation on the molecular spectral maps. We present a chart of line
formation which may serve as valuable tool for understanding results of LRT
simulations. This chart clearly demonstrates which parts of the model core
contribute to the line profile at each velocity.

We would like to make the following conclusions:
\begin{itemize}
\item Densities in starless cores fall into the range where level populations are
neither radiatively, nor collisionally dominated.
\item Large column densities do not necessarily lead to the appearance of self-absorption dips.
\item When the column density is fixed, a specific line can be optically thick only in a range of densities, being
optically thin at densities both below and above this range.
\item The density which is ``traced'' by some transition depends on external factors (UV field and CR ionization), which shape the molecular distribution.  In particular, the ``traced'' density can be lower than the critical density.
\item Rotation and infall  may produce very similar spectra and in general can only be distinguished by spectral mapping.
\end{itemize}

\acknowledgements
We are grateful to Ted Bergin, Michiel Hogerheijde, Kees Dullemond
and Juergen Steinacker for useful discussions. We also thank the referee Neal J.
Evans II, for valuable suggestions and comments.
This research has made use of NASA's Astrophysics Data System. DW, and
BS are supported by the RFBR grant 07-02-01031.

\end{document}